\documentclass[12pt,twoside]{article}
\usepackage{fleqn,espcrc1}

\usepackage{graphicx}
\usepackage[figuresright]{rotating}

\newcommand{\PL}[1]{ Phys.\ Lett.\ { #1}}

\newcommand{\PR}[1]{Phys.\ Rev.\ { #1}}
\newcommand{\PRL}[1]{ Phys.\ Rev.\ Lett.\ { #1}}

\newcommand{\AmS}{{\protect\the\textfont2
  A\kern-.1667em\lower.5ex\hbox{M}\kern-.125em}}

\title{Resonances, Chiral Symmetry, Coupled Channel Unitarity
and Effective Lagrangians
\thanks{
Talk given at the 8th International Conference on Hadron Spectroscopy, HADRON99,
August 24-28, 1999, Beijing, China.
Work partially supported by DGICYT under contracts PB96-0753 and 
AEN97-1693 and by the EU TMR network eurodaphne contract no.
ERBFMRX-CT98-0169. }}

\author{ J. R. Pel\'aez\address{ Departamento de F\'{\i}sica Te\'orica.
Universidad Complutense de Madrid\\ 28040 Madrid. SPAIN.},
J.A.Oller\address{Forschungzentrum J\"ulich, Institut f\"ur 
Kernphysik (Theorie)\\ D-52425 J\"ulich, GERMANY.}
 and E.Oset\address{Departamento de F\'{\i}sica 
Te\'orica and I.F.I.C.
Centro Mixto Universidad de Valencia-C.S.I.C.
46100 Burjassot (Valencia), SPAIN.}
}
       
\begin{document}

% typeset front matter
\maketitle

\begin{abstract}
By means of a coupled channel non-perturbative  unitary approach, 
 it is possible to extend
the strong constrains of Chiral Perturbation Theory to higher energies.
In particular, it is possible to reproduce the lowest lying resonances
in meson-meson scattering up to 1.2 GeV using
the parameters of the $O(p^2)$ and $O(p^4)$ Chiral Lagrangian.
We report on an update of these results examining their
possible relevance for meson spectroscopy.
\end{abstract}

\section{Introduction}

Chiral Perturbation Theory (ChPT) \cite{weinberg,xpt} has proved 
very successful in order to
describe the physics of mesons at very low energies. The key point
of the whole approach is to identify the lightest pseudoscalar
mesons $\pi , K$ and $\eta$ 
as the Goldstone bosons associated to the chiral symmetry
breaking. These particles will be the only degrees of 
freedom at low energies and their interactions can be described
in terms of the most general effective Lagrangian which
respects the chiral symmetry constraints. 

So far as this is a 
low energy approach, the amplitude of a given process
is basically given as an expansion in
the external momenta over the scale of symmetry breaking 
$4\pi v\simeq 1.2\,$GeV. It is also possible to calculate loops, which
increase by two the order of the term in the chiral expansion and
generate logarithmic contributions as well as divergences. The former
are very important at low energies since they can dominate
over some polynomial terms and the latter have to be absorbed 
in the renormalization of the free parameters that appear at the 
next order in the Lagrangian. It is therefore possible to obtain
results which are finite to a given order in momenta. They provide
a very good description of meson interactions up to about 500 MeV in 
the best cases.

However, if one is interested in resonances in particular, as it happens in 
meson spectroscopy, it is little what one could do with just plain ChPT.
In this work we will review recently proposed
 new nonperturbative schemes imposing unitarity to
the chiral Lagrangian, thus enlarging the convergence of 
the chiral expansion and reproducing resonances. We will briefly comment
on possible implications for meson spectroscopy.

\section{ChPT and Unitarity}

Within the coupled channel formalism, the unitarity of the $T$
matrix reads 
\begin{equation}
\hbox{Im}\,T = T\,\hbox{Im} G\,T^{*}\;\;\Rightarrow\;\;
\hbox{Im}\, T^{-1} = - \hbox{Im}\,G \;\;\Rightarrow\; \;
T=[\hbox{Re}\, T^{-1}-i\, \hbox{Im}\,G]^{-1},
\label{IAM}
\end{equation}
%(that is basically the Optical Theorem)
where $\hbox{Im}\,G$ is a {\em known} diagonal matrix, 
whose entries
are just the phase space of the intermediate states.
Indeed, $G$ is the integral of the propagators of the
two particles in the intermediate  state.
Within ChPT the amplitudes are obtained as an expansion in powers
of momenta,
i.e. $T=T_2+T_4+...$, where the subscript stands for the order 
in the expansion. Being basically a polynomial these amplitudes
 can only satisfy
{\em perturbative unitarity} and they cannot
yield poles and therefore resonances.

Within the Inverse Amplitude Method \cite{IAM,oop,ollerpaco}, the ChPT
expansion is only used for $\hbox{Re}\,T^{-1}$, which is then
used on the right hand side of eq.(\ref{IAM}). This procedure
ensures {\em exact} unitarity, while keeping the very same ChPT expansion
at low energies. Using the complete $O(p^4)$
calculation, it was first applied
to single channel $\pi\pi$ and $\pi K$ scattering 
and it was able to reproduce
several isospin and angular momentum channels and to generate dynamically
the $\sigma$, $\rho$ and $K^*$ resonances \cite{IAM}.  
In principle, 
the IAM calculations 
up to order $n$ need the complete ChPT calculations  up to the same order. 

The next step was motivated by the results obtained
when unitarizing the lowest order $O(p^2)$ ChPT scalar amplitudes 
using the Bethe-Salpeter equations (BS) \cite{BS}.
 Remarkably it was possible to fit the meson-meson scalar 
phase shifts up to 1.2 GeV and reproduce the $\sigma$,
 $f_0$ and $a_0$ resonances,
just by setting the cutoff to a natural 
value around 1 GeV, since there are no
other free parameters. The link with the IAM was established in \cite{oop}:
the BS solution is recovered from the IAM if one approximates 
$ \hbox{Re} \,T_4\simeq\,T_2\, \hbox{Re} \,G\,T_2$. 

But in order to obtain vector resonances the $O(p^2)$ is not enough. 
Unfortunately, the full $O(p^4)$  calculation is not available
for all the meson-meson scattering channels. However it was also shown in \cite{oop}
that in order to obtain a good results up to 1.2 GeV, 
it is enough to add to the BS approximation  the $O(p^4)$
tree level, that is,
\begin{equation}
\hbox{Re} T_4\simeq \underbrace{T_4^P}_{O(p^4)\, \hbox{tree}} +
\underbrace{T_2\,\hbox{Re} G\, T_2}_{\hbox{s-channel loops}}.
\end{equation}

The results are remarkable, as it can be seen in Fig.1, where we display
an updated fit to the  meson-meson phase shifts and inelasticities 
\cite{oop,phi}. Compared
with our previous work \cite{oop}, we have also
corrected a small error in one amplitude, which only has a minor effect.
\begin{figure}[t]
\begin{center}
    \includegraphics[scale=0.6]{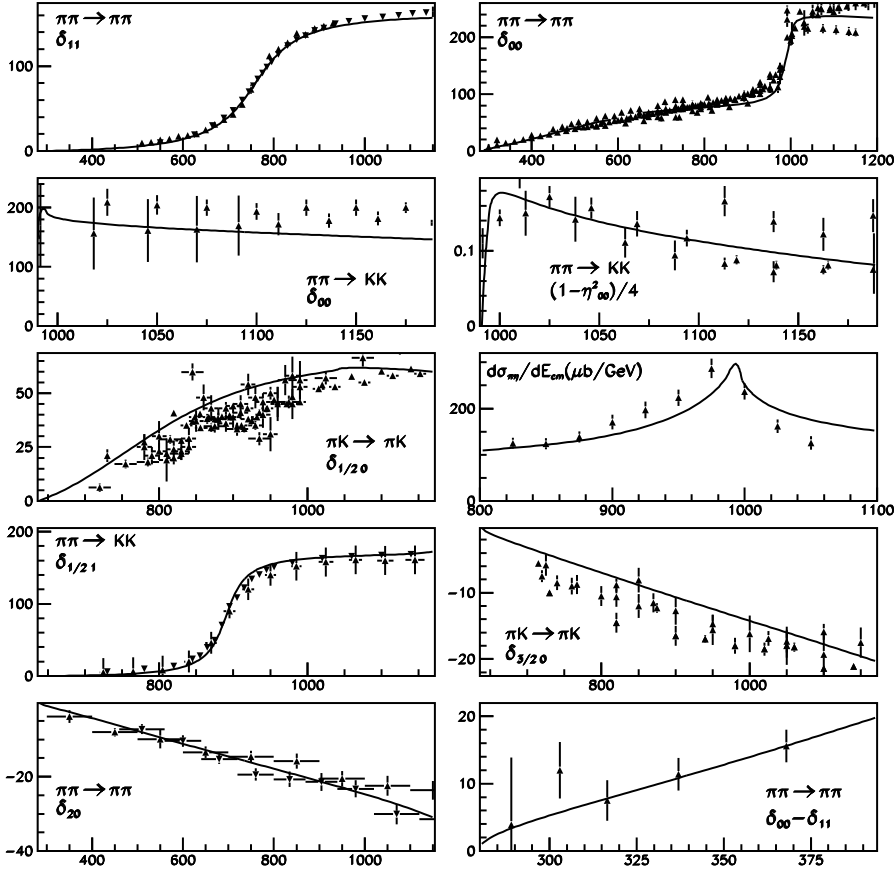}
\end{center}
\vspace{-1.5cm}
\caption{Meson-meson scattering results of the non-perturbative chiral 
approach.
For the data see references in \cite{IAM,oop,ollerpaco,oo}.}
\label{fig:largenenough}
\end{figure}
It can be noticed that we are able to reproduce together the 
$\sigma$, $f_0$, $a_0$,
$\rho$, $\kappa$ and $K^*$ resonances 
and the isospin zero state of the vector octet, $\omega_8$ ( see \cite{oop}). 
It is also possible to find, in the unphysical sheets, 
the poles associated to these resonances (see \cite{oop}).

In table I we 
list the values of the fitted IAM parameters. Note that
within this approximation, tadpoles and crossed loops are 
neglected and absorbed in the chiral parameters, so that
we cannot compare directly with the standard ChPT 
coefficients.
Nevertheless they should be of the same order of magnitude, once the 
appropriate 
renormalization scale is used. 

\begin{table}[t]
\caption{Values of the chiral constants from the IAM fit and ChPT at their equivalent scales.}
\begin{tabular}{|c|c|c|c|c|c|c|c|c|}
\hline
  & $L_1$ & $L_2$ & $L_3$ & $L_4$ & $L_5$ & $2L_6+L_8$ & $L_7$ & MeV
\\ \hline
IAM &0.88&1.54&-3.66&-0.27&1.09&0.68&0.10&$q_{max}=$673\\ \hline
ChPT&0.4$\pm$0.3&1.3$\pm$0.3&-3.5$\pm$1.1&-0.3$\pm$0.5&
1.3$\pm$0.5&0.42$\pm$0.5&-0.4$\pm$0.2&$\mu=800$ \\
\hline
\end{tabular}
\end{table}

If we had the complete $O(p^4)$ ChPT calculation for all 
the meson-meson channels
and we used  the same renormalization scheme as
in standard ChPT, the fitted parameters should be very similar to
 the standard ones. This has been checked with the single channel
$\pi\pi$ and $\pi K$ scattering, but also by calculating the complete $2\times2$
$T$ amplitude for the $\pi\pi$, $K \bar{K}$ coupled channels \cite{ollerpaco}.
In both cases the fitted parameters are perfectly compatible with those 
of standard ChPT. Further work is still in progress.

\section{Summary and Implications for Meson spectroscopy}

Chiral Perturbation Theory 
unitarized with the Inverse Amplitude Method, describes correctly the
dynamics of meson-meson scattering in coupled channels, generating:

\begin{itemize}
\item The $\rho$ and the $K^*$, as well as a pole that corresponds to the $\omega_8$,
 which, all together, form the lightest {\em vector octet}. In order
to obtain them it is necessary to include the $O(p^4)$ chiral parameters. 
The nonet is not reproduced
since in the SU(3) limit the $\omega_1$ does not couple to two mesons.
\item The $f_0$, $a_0$, $\sigma$ and $\kappa$. 
All their masses are below 1 GeV, and {\em the last two are very wide}, 
not Breit-Wigner
resonances. {\em All of them can be simply generated by unitarization 
of the lowest order ChPT}, with just a cutoff as a free parameter.
\end{itemize}
The role of the $O(p^4)$ chiral parameters
can be understood 
\cite{xpt,Ecker} by writing a Lagrangian with pions, kaons and etas, but also
with heavier resonances coupled in a chirally invariant way.  Then one integrates
out these heavier states and the resulting Lagrangian is that of ChPT, but now the
values of the chiral constants can be related to the masses and widths
of the preexisting heavier resonances (``Resonance Saturation Hypothesis''). Most of
the  experimental values of the chiral coefficients
are saturated by these estimates due to vector resonances alone 
(that is vector meson dominance) but some other parameters still need the existence
of scalar states. 
Recently  \cite{oo}, using the N/D unitarization method with
explicit resonances added to the lowest order ChPT Lagrangian,  
it has been established that these heavier scalar
states should appear with a mass around 1.3 - 1.4 GeV for the octet and 1 GeV for the singlet.
In addition, the  $\sigma$, $\kappa$, $a_0$ and a strong contribution
to the $f_0$, were also generated from the unitarization of the ChPT lowest 
order.
These states still survive when the heavier scalars are removed.
That agrees with our observation that the $\sigma$, $\kappa$, $f_0$ and $a_0$
are generated {\em independently of the chiral parameters}, that is, of the preexisting 
scalar nonet, which is heavier.

Since Chiral Perturbation Theory does not deal with quarks and
gluons, it is very hard to make any conclusive statement about the 
nature of these states
($q\bar{q}$, four-quark, molecule, etc...), unless we make additional assumptions. 
However, it seems clear that the nature of the lightest
scalar mesons is different from that of vectors. In addition, the fact that we obtain
simultaneously the above nine scalar resonances with the same procedure, seems to
indicate that they are good candidates to form a complete $SU(3)$ nonet.

\end{document}